\newcommand{\beq}{\begin{eqnarray}}
\newcommand{\eeq}{\end{eqnarray}}

\documentstyle[prl,twocolumn,aps]{revtex}
\begin{document}
%%\draft

\narrowtext

\noindent

\noindent
{\large \bf Incipient p-wave Superconductivity in a Si MOSFET}

\vskip 3mm

In a recent letter, Kravchenko and colleagues\cite{krav} reported a
surprising insulator-metal transition (IMT)
in a 2D high-mobility silicon metal-oxide-semiconductor field-effect
transistor
(MOSFET) as the electron density, $n_s$, increased from $8.14\times
10^{10}/cm^2$ to $8.8\times 10^{10}/cm^2$.
Two key features that characterize the ``IMT''
are 1) the symmetry of the I-V curves around the
I-V curve at the transition density, $n_c=8.47\times 10^{10}/cm^2$
and 2) the scaling of the resistivity as a function of
electric field and temperature:
$\rho(T,n_s)=f_1(|\delta_n|/T^b)$ with $b=1/z\nu$
and
$\rho(E,n_s)=f_2(|\delta_n|/E^a)$, $a=1/[(z+1)\nu]$,
$\delta_n=(n_s-n_c)/n_c$, $\nu=1.5$ and
$z\approx 0.8$. The fact that $z\approx 1$ signifies
that electron correlations are dominant.
Interpreted\cite{krav} as a true IMT, these
experiments suggest that electron
correlations thwart localization in 2D disordered systems.
Theoretically, however, this question is still open.

We argue here that the observed ``IMT''\cite{krav}
is consistent instead with a transition
from an insulator to a T=0 p-wave superconductor.
Hence, the yet-unproven metallic state in 2D does not have to be invoked
to explain the experimental data.
Our argument consists of three parts.  Consider first the
I-V curves.  In the insulator, the I-V curves indicate that a threshold
voltage
must be applied before current flows.  This suggests that
transport is
activated in the insulator.  Experimentally\cite{krav2},
the activation energy, $\Delta$, was measured to vanish at the
transition
density\cite{krav2}.
In the conductor, the I-V curves exhibit negative
curvature in the vicinity of zero applied voltage.  This signifies a
crossover
to
a state in which current flows in the absence of an applied voltage
as in a superconductor.  That this state of affairs obtains is simply a
consequence
of the symmetry between I and V on either side of the
transition\cite{shah}.  If I
and V can
be
interchanged, then a charge activation gap in the insulator will manifest
itself as
a supercurrent on the conducting
side. Naively, the magnitude of the supercurrent is proportional to the 
superconducting gap. We conjecture that 
$I\Leftrightarrow V$ symmetry might be related
to the dual role played by the gap across the transition.
At the transition density,
the single particle localization length\cite{malee} is equal to
the Cooper pair coherence length.
While single electrons are localized by disorder, Cooper pairs at the
brink of
superconductivity
can diffuse giving rise to a linear I-V characteristic\cite{goldman}.
Next, the electric field and temperature scaling of the resistivity
given above is a natural consequence of the insulator-superconductor
transition (IST) \cite{ali}. As a $T=0$ transition, however, only superconducting phase or pair fluctuations
will occur on the conducting side.

What about the pairing mechanism? At densities of $10^{11}/cm^2$, the
electron
interaction, $V_{e}\approx 10\epsilon_F$\cite{krav}.  Such strong
correlations render standard BCS s-wave pairing mechanisms ineffective
in driving the IST.  However, at such low densities\cite{fowler}, 
we can expand the potential
in powers of $q/q_s$ for $q\approx 2p_F$ because $p_Fq_s^{-1}\ll 1$.
Here
$q_s^{-1}=\lambda_s$ is a reciprocal screening length which for the 2D
electron gas is of the order of the Bohr radius\cite{fowler}.  
Consequently, $V_e(q=0)- V_e(2p_F)\approx 0$
and the s-wave component in the potential dominates.  In this limit,
Chubukov\cite{chub} showed
for an arbitrary short-range repulsive
interaction, $p_Fa\ll 1$, where $a$ is the scattering length,
the dominant vertex renormalization of the scattering function
$\Gamma(q)$ arises from third-order ladder exchange diagrams
in the particle-particle channel.  For $q\le 2p_F$ such renormalizations
lead
to a p-wave interaction that is negative and as a consequence
triplet ($S=\pm 1$) p-wave superconductivity at T=0.  A curious feature of
the exchange
mechanism is that it is destroyed predominantly by polarizing the spins
in the intermediate scattering states.
Hence, if the spins are polarized
as in either a parallel or
perpendicular magnetic field, the third-order vertex correction
vanishes as should the p-wave pairing interaction.
Recent
experiments\cite{krav3} indicate that the conducting state is indeed
suppressed
by either a parallel or a perpendicular magnetic field,
consistent with a spin-polarization mechanism for the destruction of
the superconducting state. 
Pairing fluctuations
should vanish when the spin energy $g\mu_B H\approx\epsilon_F$; hence 
the characteristic field $H_c$ should increase
as a linear function of doping, $\delta_n$. For $\epsilon_F=0.6{\rm mev}$, $H_c\approx 3T$.
Experimentally, $H_c\approx 1T$ and increases with $\delta_n$\cite{krav3}.  Further susceptibility and phase 
sensitive measurements are needed to confirm the p-wave scenario presented here.       

We acknowledge useful conversations with 
E. Fradkin, who sparked our interest in the possibility of superconductivity, 
I. Martin, S. Kravchenko, A. Chubukov, 
N. Ashcroft, A. Castro-Neto, and P. Parris and the NSF 
grant No. DMR94-96134.

\vskip 3mm
\noindent Philip Phillips and Yi Wan

{\small Loomis Laboratory of Physics, University of Illinois  \par
1100 W.Green St., Urbana, IL, 61801-3080 }


\begin{thebibliography}{99}

\bibitem{krav}
S. V. Kravchenko, et. al.
Phys.
Rev. Lett. {\bf 77}, 4938 (1996).
\bibitem{krav2}
V. M. Pudalov, et. al. Phys. Rev. Lett., {\bf 70}, 1866 (1993).
\bibitem{shah}
D. Shahar, et. al.
Phys. Rev. Lett. {\bf 74}, 4511 (1995).
\bibitem{malee}
M. Ma and P. A. Lee, Phys. Rev. B {\bf 32}, 5658 (1985).
\bibitem{goldman}
M. P. A. Fisher, Phys. Rev. Lett. {\bf 65}, 923 (1990); Y. Liu and A. M. Goldman, Mod.
Phys. Lett. B {\bf 8}, 277 (1994).
\bibitem{ali}
A. Yazdani and A. Kapitulnik, Phys.
Rev. Lett. {\bf 74}, 3037 (1995)
\bibitem{fowler}T. Ando, et. al. Rev. Mod. Phys. {\bf 54}, 437 (1982).
\bibitem{chub}
A. Chubukov, Phys. Rev. B {\bf 48}, 1097 (1993).
\bibitem{krav3}
D. Simonian, S. V. Kravchenko, and M. P. Sarachik, unpublished data, 1997.
\end{thebibliography}
\end{document}